\documentclass[twocolumn,showpacs,superscriptaddress,prl,amsmath,amssymb,letterpaper]{revtex4-1}
\usepackage{times}
\usepackage{amsmath,bm,amsfonts}
\usepackage{dcolumn}
\usepackage{graphicx}
\usepackage{latexsym}

\begin{document}

\title{Gauge Field and Confinement-Deconfinement 
Transition in Hydrogen-bonded ferroelectrics}
\author{Chyh-Hong Chern}
\affiliation{Department of Physics, 
National Taiwan University, Taipei 10617, Taiwan}
\author{Naoto Nagaosa}
\affiliation{Department of Applied Physics, University of Tokyo, 
Tokyo 113-8656, Japan}
\affiliation{RIKEN Center for Emergent Matter Science (CEMS), Wako 351-0198, Japan}
\date{\today}

\begin{abstract}
Quantum melting of ferroelectric moment in the frustrated hydrogen-bonded 
system with "ice rule" is studied theoretically by using the quantum Monte Carlo 
simulation. The large number of nearly degenerate configurations are described as
the gauge degrees of freedom, i.e., the model is mapped to a
lattice gauge theory which shows the confinement-deconfinment transition (CDT).
The dipole-dipole interaction $J_2$, on the other hand, explicitly breaks
the gauge symmetry leading to the ferroelectric transition (FT) at finite temperature $T$. 
It is found that the crossover from FT to CDT manifests itself in the 
reduced correlation length of the polarization 
$\xi_{\text{FT}} \sim \Delta (K-K_c)^{-\nu}$ with
$\Delta \propto \sqrt{J_2}$ while $K_c$ and $\nu$ remains finite 
in the limit $J_2 \to 0$. In contrast, the Currie-Weiss-like law for the susceptibility $\chi$ 
and the spontaneous polarization 
behaves smoothly and the length scale $\xi_{\text{CDT}}$, related to the molecular symmetry and volume for CDT, does not reduce in this limit. 
\end{abstract}

\maketitle

  The hydrogen-bonded systems are one of the 
ideal laboratories to study the quantum tunneling. Especially, the
ferroelectric properties of these systems attract much attention 
since the old work by Slater on KH$_2$PO$_4$ (KDP)~\cite{Slater}. The quantum melting of 
the ferroelectric order to result in the quantum paraelectricity
is a rather common phenomenon observed in several hydrogen-bonded 
ferroelectrics~\cite{quantum1,quantum2,quantum3,SQacid}, 
which is usually described by the transverse Ising model
\begin{equation}
H  = - \sum_{ij} J_{ij} \sigma^z_i \sigma^z_j - K \sum_i \sigma^x,
\label{eq:TIsing}
\end{equation}
where $\sigma^z = \pm 1$ specify the positions of the 
hydrogen atoms, $J_{ij}$ is the dipole-dipole interaction, and 
$K$ represents the tunnelling matrix element. These two interactions
compete with each other, and by increasing $K$,
there occurs a phase transition from the ordered state to the quantum disordered
phase.

  On the other hand, it often happens that the 
constraints are significant to the hydrogen-bonded systems.
Actually, the hydrogen positions in the 
representative system KDP are already subject
to the constraint, i.e., so called "ice rule"~\cite{Slater}. 
Namely, only two of the four hydrogen atoms 
next to a tetrahedron are approaching to the
center for the low energy sector. 
Similar constraint is also relevant to the recently studied 
quasi-two dimensional antiferroelectric squaric acid (H$_2$SQ), where 
the square molecule is surrounded by 4 molecules with 
hydrogen bonds~\cite{SQacid}, and the two-in-two-out configurations are
energetically stable. 
This  "ice rule" is the generalization of the 
hydrogen bonds in ice leading to the macroscopic 
degeneracy of the ground state configurations as 
discussed by Pauling long time ago~\cite{Pauling}. 
Therefore, a keen issue is how this macroscopic
degeneracy of the low-energy states in the hydrogen bonded systems affects the nature of the phase transition.
   
The constraints imposed on the physical variables are 
more common phenomenon found in many other 
cases. Frustrated magnets are one of such examples, 
where some of the macroscopically degenerate spin configurations 
are selected as the lowest energy states. 
Spin ice in pyrochlore ferromagnet is a representative
example, in which the hydrogen position is replaced by
the direction of the spin, and the "ice rule" applies
simultaneously in every tetrahedron.  This property leads to an interesting 
phenomena, e.g., the absence of the long range ordering
down to zero temperature and the deconfined magnetic monopoles   
as the excitations~\cite{Castelnovo}. These are described well by the gauge theory 
representing the constraints within the framework of 
the classical statistical mechanics. Quantum effects
on the spin ice model have been attracting intense
interests recently~\cite{Ross, Shannon}.
\begin{figure}[htb]
\includegraphics[width=0.45\textwidth]{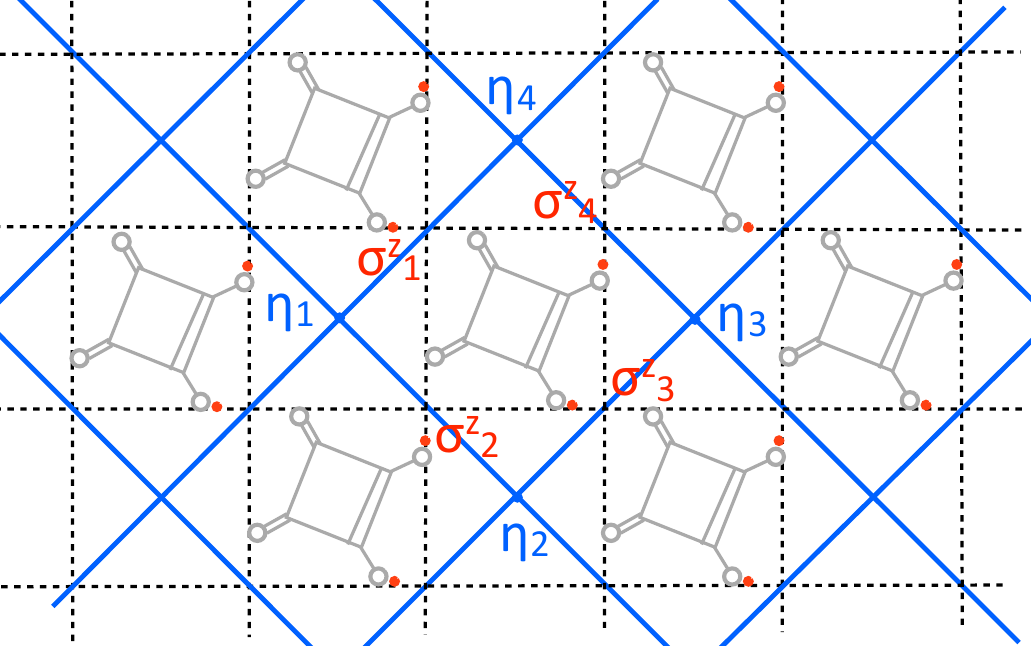}
\caption{(color online) The two-dimensional square lattice (blue lines) formed by the squaric-acid molecules.  The red balls label the hydrogen ions.  The $\eta$-variables are defined on the lattice sites.  The hydrogen positions are parametrized by the $\sigma^z$-variables, which are defined on the lattice bonds forming a two-dimensional dual lattice (black dash lines).}\label{Fig:lattice}
\end{figure}

In this Letter, we develop a theory for the organic ferroelectrics with macroscopic degeneracy.  A $Z_2$-gauge-invariant term accounting for the "ice rule" is introduced explicitly~\cite{Maier}.  
Different from a $U(1)$ gauge theory, our model exhibits two types of quantum phase transitions, i.e., the
confinement-deconfinement transition (CDT) of the gauge field and the
ferroelectric transition (FT) of the local dipole moments.  We relate these two phenomena by introducing a dipole-dipole interaction $J_2$ explicitly breaking the gauge symmetry (Eq.(\ref{h2}) below).  Due to the macroscopic degeneracy, different from the ordinary FT, we found two length scales $\xi_{\text{FT}}$ and $\xi_{\text{CDT}}$ (defined later) in the vicinity of the FT as the system is close to the CDT.

\begin{figure}[htb]
\includegraphics[width=0.45\textwidth]{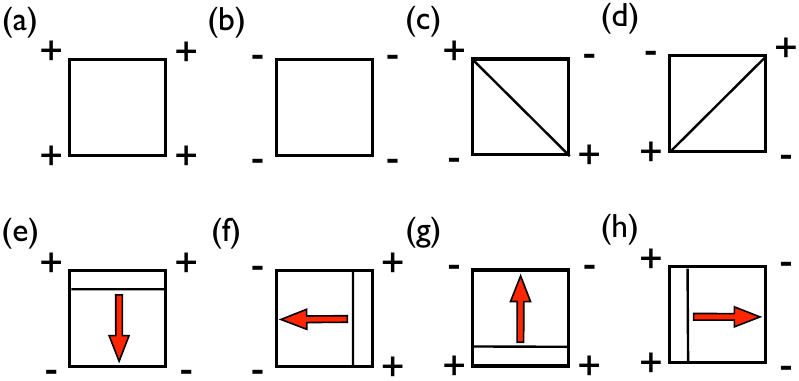}
\caption{(color online) The ground state configuration of each plaquette in the model of $H_0$ in Eq.~(\ref{h0}).  There are finite dipole moments in the states (e) to (h).  The directions of the dipole moment are drawn in red for the molecules of group A.}\label{Fig:molecule}
\end{figure}

Taking the squaric acid as a prototype, we consider a two-dimensional model, 
$H = H_0 + H_1 + H_2$,
where
\begin{eqnarray}
\!\!\!\!&H_0& = -J_0\sum_{\Box}\sigma^z_1\sigma^z_2\sigma^z_3\sigma^z_4-K\sum_{i}\sigma^x_i, \label{h0}\\
\!\!\!\!&H_1& = J_1\sum_{\Box}(\sigma^z_1\sigma^z_3+\sigma^z_2\sigma^z_4), \label{h1} \
H_2=-J_2\!\sum_{<AB>}\!\vec{P}_A\cdot\vec{P}_B \label{h2}
\end{eqnarray}
in the lattice in Fig.~\ref{Fig:lattice}, where the summation of $\Box$ in 
Eq.~(\ref{h0}) and (\ref{h1}) is over the plaquetts of the blue lattice in Fig.~\ref{Fig:lattice} resembling the H$_2$SQ molecules and $\sigma$-variables are defined on the bonds of the plaquetts~\cite{SQacid}.  On each lattice bond, there is a hydrogen ion shared by two neighboring molecules, representing the hydrogen bond.  We use $\sigma^z$ to parametrize 
the position of hydrogen ions in the following way: If a hydrogen is closer to the molecule A, it is the "$+$" state, otherwise it is a "$-$" state, representing a gauge field.  The $H_2$ in Eq.~(\ref{h2}) represents the nearest-neighbor dipole-dipole interaction, and the components of the dipole moment $\vec{P}_i$ are defined by $P_{(A, B)x}=(\pm)\frac{1}{4}(\sigma^z_1+\sigma^z_2-\sigma^z_3-\sigma^z_4)$ and $
P_{(A, B)y}=(\pm)\frac{1}{4}(\sigma^z_2+\sigma^z_3-\sigma^z_1-\sigma^z_4)$, where $(+)$
for the molecule A and $(-)$ for the molecule B respectively. 

The "ice-rule" constrained by the gauge term $J_0$ and the Ising term $J_1$ generates a macroscopic degeneracy, which is distinct from one  in the antiferromagnetic Ising model in the 2D pyrochlore (checkerboard) lattice~\cite{Shannon1} and the quantum vertex model~\cite{Olav, Ardonne}.  The gauge term favors 8 different configurations in the low energy sector illustrated in Fig.~\ref{Fig:molecule}.  Note that this quantum Hamiltonian $H_0$ corresponds to the (2+1)-dimensional Ising gauge theory in the temporal gauge, i.e., the time-component of the gauge field is fixed to be one.  The addition of $J_1$ term lifts the degeneracy so that the states of (e) to (h) are remained.  They are particularly interesting because they carry finite dipole moments.  For example, the direction of the dipole moments for the molecule A are shown in red arrows in Fig.~\ref{Fig:molecule}.  

\begin{figure}[htb]
\includegraphics[width=0.4\textwidth]{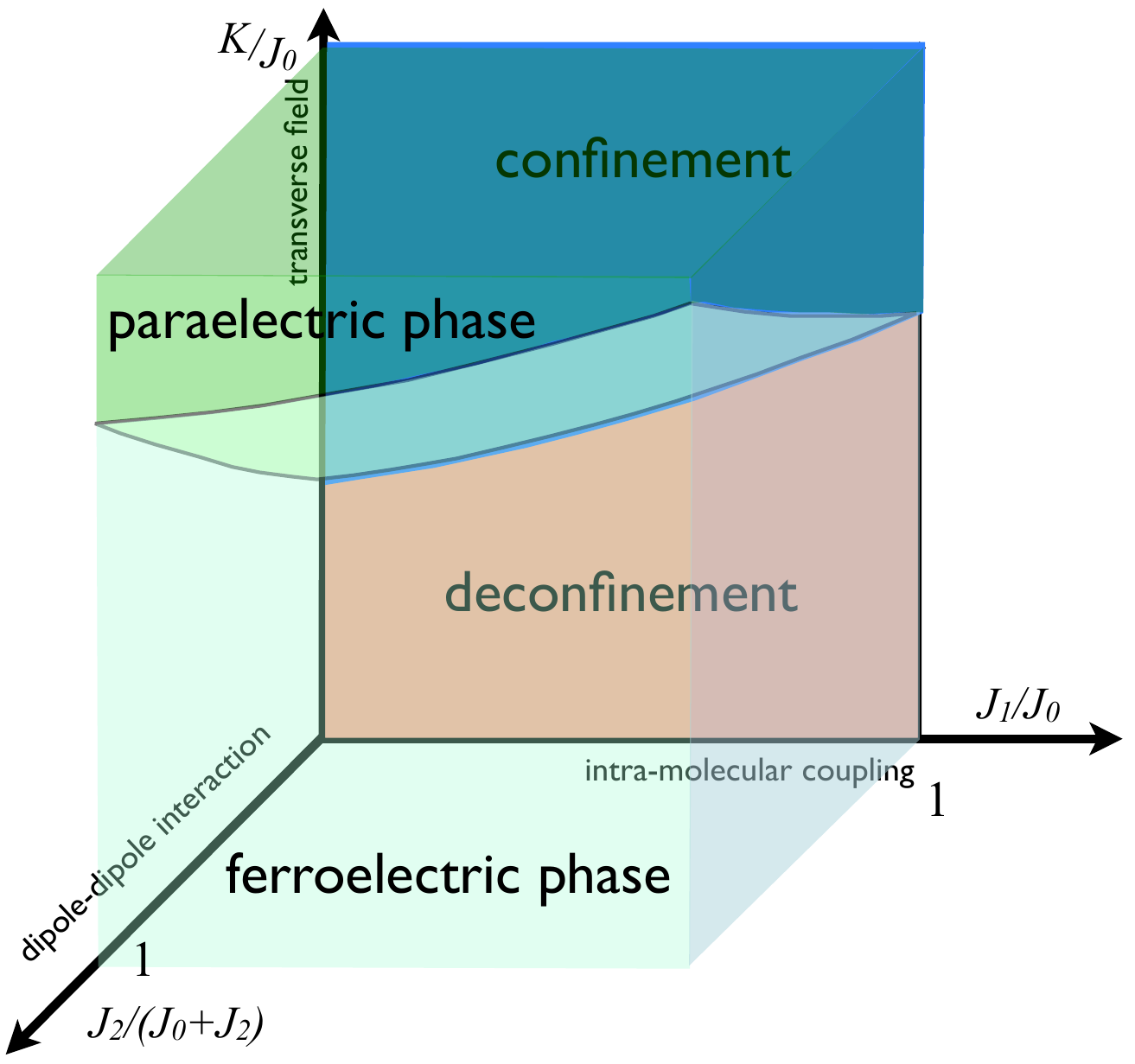}
\caption{(color online) The zero-temperature phase diagram.  For $J_2=0$, there are two phases separated by a confinement-deconfinement phase transition.  For finite $J_2$, the phase space is divided by a second-order ferroelectric phase transition.}\label{Fig:QPD}
\end{figure}

The finite-temperature property due to the gauge term was studied preciously by Maier, et al.~\cite{Maier}.  Here, we focus on the quantum phase transition in the presence of the transverse field.  The quantum phase diagram at zero temperature can be summarized in Fig.~\ref{Fig:QPD}.  When $J_1=J_2=0$, there is a second-order confinement-deconfinement transition (CDT) at critical $K_c$~\cite{Kogut, Savit}.  At the first glance, the introduction of the $J_1$ term breaks the gauge symmetry and the CDT.  However, there remains a hidden gauge symmetry.  To see this, one can introduce the $\eta$ variables defined in the dual lattice in Fig.~\ref{Fig:lattice}.  Redefining the $\sigma^z$-variable as $\sigma_j^z=\eta_i\eta_j$ in a restricted Hilbert space of the minimum $J_0$ energy, for $J_2=0$, we obtain the action
\begin{eqnarray}
S =-\beta J_0 + \frac{2\beta J_1}{n}\sum_{\Box}\eta_i\eta_j\eta_k\eta_l\!-\!K'\sum_{\Box'}\eta_i\eta_j\eta_{i'}\eta_{j'}, \label{eta}
\end{eqnarray}
where $\Box$ ($\Box'$) are the plaquettes in the spatial (imaginary-time) direction in the dual lattice, $n$ is the dimension in the imaginary-time direction, and $\beta=1/(k_B T)$. In Eq.~(\ref{eta}), we express the $\sigma^x$-term in the $\eta$-variables with $K'=\log(\coth(\beta K/n))/2$.  The hidden symmetry protects the CDT to extend to the finite $J_1$ region.  We also perform the quantum Monte Carlo calculation to confirm this.  The numerical results are prepared in the Supplementary Information~\cite{SI}.  Our analysis indicates that the CDT is a robust transition, distributing over a wide range in the phase diagram where the ice rule is satisfied.  As a first result, the phase diagram is divided into the deconfined phase and the confined phase in the $J_2=0$ plane.

\begin{figure}[htb]
\includegraphics[width=0.48\textwidth]{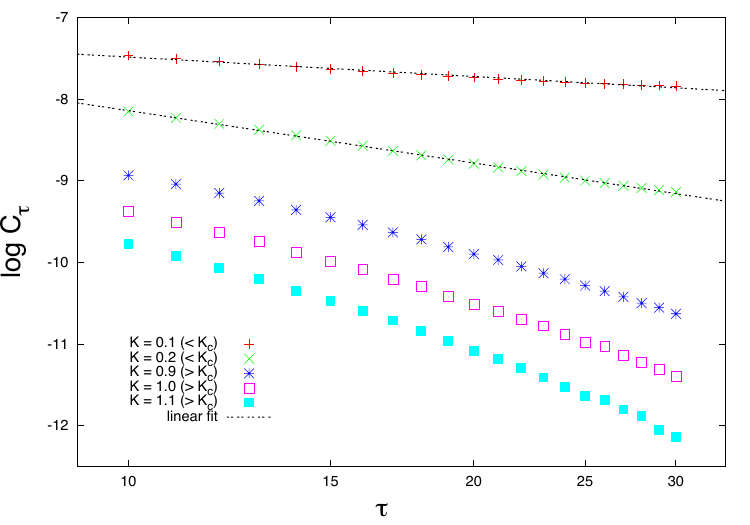}
\caption{(color online) The logarithmic plot of the correlation in the imaginary-time direction at $J_2=0$, where CDT takes place. $\tau$ is the coordinate in the imaginary-time direction.  Using $J_1 = 0.2 J$, the $K_c$ of CDT at $J_2=0$ is 0.64 as shown in the inset of Fig.~\ref{Fig:dieX}.  We plot the results for $K = 0.1$ and $0.2$ ($< K_c$) to demonstrate the power-law behaviour in the deconfined phase and $K = 0.9$, 1.0, and 1.1 ($> K_c$) to show the exponential decay in the confined phase. Note that $C_\tau$ is independent of $\tau$ for $K=0$, and the numerical error is $0.1\%\sim1\%$.}
\label{Fig:CI}
\end{figure}

Even without the dipolar interaction $J_2$, the dielectric susceptibility in the novel deconfined phase diverges for $K \le K_c$ at $T=0$.  In Fig.~\ref{Fig:CI}, we perform the Monte Carlo calculation to compute the correlation in the imaginary-time direction, defined by
\begin{eqnarray}
C_{\tau} = \frac{1}{N}\sum_i <P_x(i,\tau)P_x(i,0)>, \label{CI}
\end{eqnarray}
where $\tau$ is the coordinate in the imaginary-time direction.  The temperature $k_B T=0.02 J_0$ and the range of $\tau$ is $10< \tau< 30$ in Fig.~\ref{Fig:CI}.  The Monte Carlo simulation is performed in the lattice up to $32\times 32$ sites in $10^6$ Monte Carlo steps.  The details of the Monte Carlo simulations are given in the Supplementary Information.   Under the temporal gauge, Eq.~(\ref{CI}) contains gauge-invariant terms; i.e., $<\sigma_i (\tau) \sigma_i(0)> \neq 0$.  Thus, $C_\tau$ does not vanish due to the gauge symmetry at $J_2=0$.   We obtain that $C_\tau$ has a power-law decay for $K < K_c$ and $C_\tau$ has an exponential decay for $K > K_c$.  Therefore, the dielectric susceptibility,
\begin{eqnarray}
\chi = \frac{1}{N}\sum_{i,j}\int_0^{\beta}d\tau<P_x(i, \tau)P_x(j,0)>, \label{dieX}
\end{eqnarray}
diverges for $K \le K_c$ at $T=0$.  We note that at $K=0$ the system is classical and $\chi \sim 1/T$, also diverging at $T=0$. 

\begin{figure}[htb]
\includegraphics[width=0.5\textwidth]{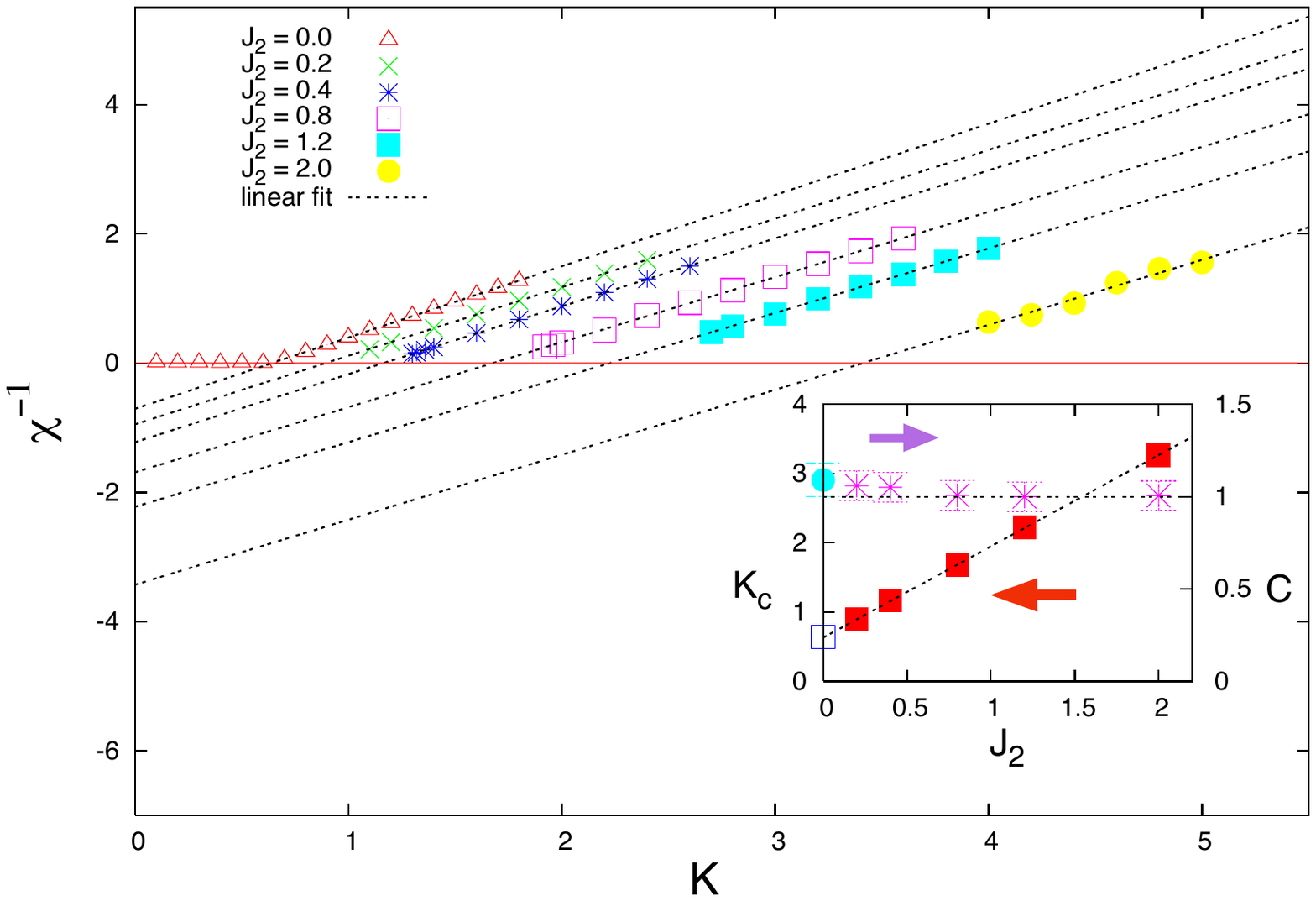}
\caption{(color online) The inverse of the dielectric susceptibility $\chi^{-1}$ for various $J_2$ values is computed for $J_0=1$ and $J_1=0.2$ at $T=0.05J_0$.  The dielectric susceptibility satisfies well the Curie-Weiss-like behaviour $\chi= C/(K-K_c)$ for $K > K_c$. We plot, in particular, the full range of $K$ for the $J_2=0$ case to show the divergence of the susceptibility for $K<K_c$.  In the inset, we extract $C$ and $K_c$ as functions of $J_2$. The data of $C$ uses the right $y$-axis and $K_c$ uses the left one.  $C$ remains unity for all $J_2$.  $K_c$ has a linear relation with $J_2$, terminating at $K_c = 0.64$ for $J_2=0$, where the confinement-deconfinement phase transition takes place. Note that the numerical error is $\sim$0.2\%.}
\label{Fig:dieX}
\end{figure}

Introducing the dipolar interaction $J_2\neq 0$, the ground state develops a spontaneous polarization 
for $K<K_c(J_2)$.  At finite temperature, a ferroelectric transition can occur.  Correspondingly, $\chi$ diverges at the critical temperature $T_c$ and the power-law behavior of $\chi$ as $T \to 0$  disappears.  Moreover, a quantum phase transition to the dielectric state can be driven by increasing $K$.  
In Fig.~\ref{Fig:dieX}, $\chi$ is computed for 6 different $J_2$ values.  The dielectric susceptibility satisfies the Curie-Weiss-like behaviour $\chi=C(K-K_c)^{-1}$ for \emph{all} $J_2$ and $K_c$ vary with $J_2$ as shown in the inset, which establishes our first relation between CDT and FT.  The confined phase at $J_2=0$ and the dielectric phase for finite $J_2$ share the similar $K$-dependence.  $C$, shown in the inset of Fig.~\ref{Fig:dieX}, are \emph{independent of $J_2$} and $K_c$ of FT converges to a finite value, indicating that the FT is robust and the dipolar interaction $J_2$ is a relevant perturbation.  The convergent value of $K_c$ at $J_2=0$ is the one for CDT taking place.  Extending to the finite $J_2$ region, the 
deconfined phase at the zero-$J_2$ plane becomes the ferroelectric phase, and the confined 
phase becomes the dielectric phase as depicted in Fig.~\ref{Fig:QPD}.  Due to these non-trivial 
connections, how does the criticality of the confinement-deconfinement phase transition (CDT) of the 
gauge field affect the criticality of the ferroelectric phase transition (FT) is the main scope of this Letter.

\begin{figure}[htb]
\includegraphics[width=0.48\textwidth]{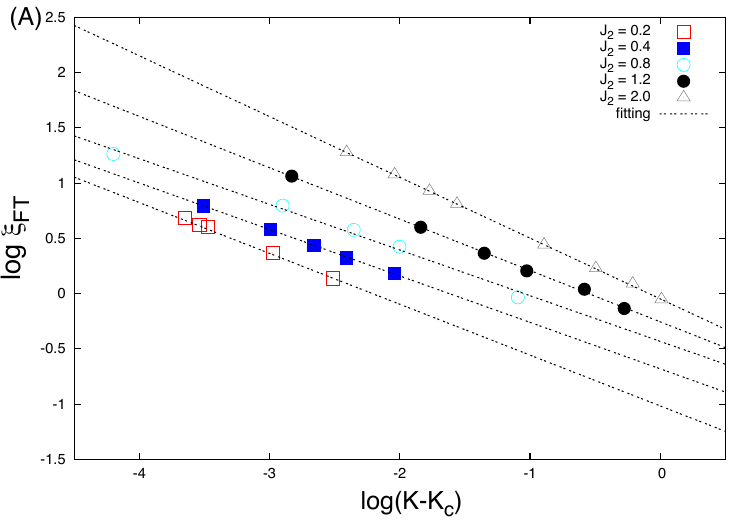}
\includegraphics[width=0.48\textwidth]{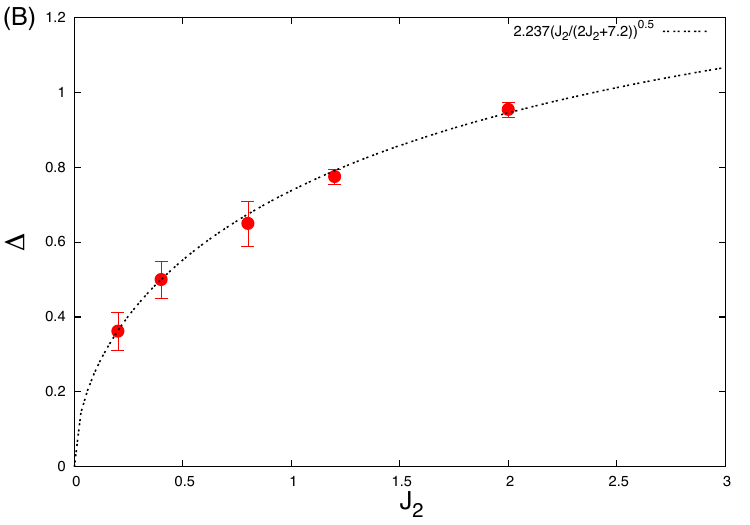}
\caption{(color online) (A) The logarithmic plot of the correlation length $\xi_{\text{FT}}$.  $\nu$ are obtained as $0.55$, $0.46$, $0.41$, $0.42$, $0.46$ for $J_2$ = $2.0$, $1.2$, $0.8$, $0.4$, and $0.2$ respectively.
 (B) The coefficient $\Delta$ as the function of $J_2$.  
The dot curve is the fitting by the mean-field result.}
\label{Fig:CF}
\end{figure}

To answer this question, we compute the ferroelectric correlation length $\xi_{\text{FT}}$, defined by
\begin{eqnarray}
<P_s(r)P_s(0)>\ \ \propto \ \ \frac{e^{-r/\xi_{\text{FT}}}}{r}, \ \ s = x \ \ \text{or}\ \ y, \label{correlation}
\end{eqnarray}
for $K>K_c$ in Fig.~\ref{Fig:CF}. The FT is a second-order phase transition because both dielectric constant and the correlation length diverge at $K=K_c$.  As shown in Fig.~\ref{Fig:CF}A, $\xi_{\text{FT}}$ obeys $\sim \frac{\Delta}{(K-K_c)^\nu}$ very nicely with $\nu$ = $0.46$, $0.42$, $0.41$, $0.46$, $0.55$ for 
$J_2 = 0.2, 0.4, 0.8, 1.2, 2.0$.  We believe that the fluctuation 
of the $\nu$ comes from error bars in the estimation.  Those values of $\nu$ are closer to the mean-field values from the 3D Ising value $\nu \approx 0.6$, indicating that the systems are outside the critical region. For a conventional ferroelectric system without macroscopic ground-state degeneracy, $\Delta$ is a constant independent of the coupling constant $J_2$.  However, we obtain $\Delta\sim\sqrt{J_2}$ as $J_2 \to 0$ in the numerical calculations and in the mean-field theory (detailed in the Supplementary Information) as shown in Fig.~\ref{Fig:CF}B.  The connection between the FT and the CDT is highly non-trivial and can be understood as the following.  As well known, any physical quantity without gauge invariance has zero ground-state expectation value in the gauge-invariant theory~\cite{Kogut, Savit}.  The correlation function at finite distance in Eq.~(\ref{correlation}) should be zero at $J_2 = 0$, since the $<\sigma^z_i\sigma^z_j>$  for $i\ne j$ therein is not gauge invariant with respect to the spatial gauge transformation.   Consequently, although the polarization moment $C$ remains unity at $J_2=0$, not only the system does not order but also the spatial correlation is restricted to zero.  In other words. the dipolar interaction in Eq.(\ref{h2}) introduces nothing but the $k$-dependence of the ferroelectric wave. When $J_2=0$, the system is free of spatial coupling and therefore $\xi_{\text{FT}}$ vanishes.  This profound feature provides a good measure of distance for a ferroelectric system in the vicinity of the CDT.  The measurement of the spatial correlation length by neutron scattering or Raman scattering~\cite{Okimoto,reiter} toward the phase transition can be used to detect whether or the system is near the CDT.

Bordered by the deconfined phase, the ferroelectric phase for small but finite $J_2$ is different from the conventional ferroelectric materials.  The frustration due to the "ice rule" is constrained by the molecular symmetry and volume, which also extends to finite $J_2$.  At $J_2=0$, it can be parametrized by the product of four 
$\sigma^z$'s in a molecule, i.e., $p_i = \sigma^z_1\sigma^z_2\sigma^z_3\sigma^z_4$.  In the critical region, the correlation $<p_i p_j >$ is proportional to $R^{-(d - \alpha/\nu)} g( R/\xi_{\text{CDT}})$, where $R= |R_i- R_j|$ is the distance between molecules $i$ and $j$, while $\alpha$ and $\nu$ 
are the critical exponent for the specific heat and the correlation length of the corresponding 3D Ising model, respectively.  The function $g$ is a scaling function and $\xi_{\text{CDT}}$ is the correlation length in the 3D Ising model, which diverges at $K=K_c$~\cite{Savit}.  The correlation naturally extends to finite $J_2$ region.  Therefore, there are two length scales behaving differently in the ferroelectric phase in the small $J_2$ region.  As $\xi_{\text{FT}}$ converges to the atomic scale, $\xi_{\text{CDT}}$ remains finite at $J_2=0$.  Representing the molecular symmetry, $\xi_{\text{CDT}}$ can actually be measured in the non-resonance Raman scattering as discussed in the Supplementary Information.

In conclusion, the effect of the ice rule and consequent gauge symmetry in the hydrogen-bonded ferroelectrics is intricate.  Although the confinement-deconfinement transition at $J_2=0$ cannot be described by the local order parameter, it can be indirectly probed by the measurement of the dielectric susceptibility.  For $K > K_c$, the system is in the confined phase with the dielectric susceptibility obeying a Currie-Weiss-like law.  For $K \le K_c$ the system is in the deconfined phase with a divergent dielectric susceptibility.  As soon as the dipolar interaction $J_2$ is turned on, ferroelectric phase develops for $K \le K_c$ as $T$ is lowered.  When the dipolar interaction is small, the approximate gauge invariance suppresses the growth of the critical region by regulating the spatial correlation length $\xi_{\text{FT}}$ obeying $\sim \Delta (K-K_c)^{-\nu}$.  We demonstrate $\Delta \sim \sqrt{J_2}$ both in the numerical simulation and in the mean-field treatment.  Our theory provides a scheme to uncover the shadow of the gauge field as well as to realise the accompanying CDT by identifying the two length scales $\xi_{\text{FT}}$ and $\xi_{\text{CDT}}$ near the ferroelectric phase transition.  A future research direction can be a further extension to include the long-ranged dipolar interaction.  Most importantly, a theory to describe the class of FT belonging to the first-order phase transition should be developed.

The authors acknowledge the fruitful discussion with Y. Tokura. 
This work is supported by Grant-in-Aid for Scientific Research
(Grants No. 24224009) from the Ministry of Education, Culture,
Sports, Science and Technology of Japan, Strategic International Cooperative Program (Joint Research Type) 
from Japan Science and Technology Agency, and Funding
Program for World-Leading Innovative RD on Science and Technology (FIRST Program) (NN).  
It is also supported by National Science Council of Taiwan under the grant: 
NSC 100-2112-M-002-015-MY3 (CHC).  CHC is grateful for the travelling support from 
Center for Theoretical Sciences in NTU.

\end{document}